\documentclass[twocolumn,amsmath,amssymb,aps,prb]{revtex4-2}
\usepackage{graphicx}
\usepackage{hyperref}
\usepackage{xcolor}
\usepackage{multirow}
\begin{document}
	
	\preprint{APS/123-QED}
	
	\title{Electrical and thermal transport through $\alpha -T_3$ NIS junction} 

\author{Mijanur Islam$^{1}$}
\author{Priyadarshini Kapri$^2$}

\affiliation{$^1$Department of Physics, Indian Institute of Technology-Guwahati, Guwahati-781039, India.\\
	$^2$Department of Physics, Osaka University, Osaka 560-0043, Japan.}


\begin{abstract}
We investigate the electrical and thermal transport properties of the $\alpha-T_3$ based normal metal-insulator-superconductor (NIS) junction using Blonder-Tinkham-Klapwijk (BTK) theory. We show that the tunneling conductance of the NIS junction is an oscillatory function of the effective barrier potential ($\chi$) of the insulating region upto a thin barrier limit. The periodicity and the amplitudes of the oscillations largely depend on the values of $\alpha$ and the gate voltage of the superconducting region, namely, $U_0$. Further, the periodicity of the oscillation changes from $\pi$ to $\pi/2$ as we increase $U_0$. To assess the thermoelectric performance of such a junction, we have computed the Seebeck coefficient, the thermoelectric figure of merit, maximum power output, efficiency at the maximum output power of the system, and the thermoelectric cooling of the NIS junction as a self-cooling device. Our results on the thermoelectric cooling indicate practical realizability and usefulness for using our system as efficient cooling detectors, sensors, etc., and hence could be crucial to the experimental success of the thermoelectric applications of such junction devices. Furthermore, for an $\alpha-T_3$ lattice, whose limiting cases denote a graphene or a dice lattice, it is interesting to ascertain which one is more suitable as a thermoelectric device and the answer seems to depend on the $U_0$. We observe that for an $\alpha-T_3$ lattice corresponding to $U_0=0$, graphene ($\alpha=0$) is more feasible for constructing a thermoelectric device, whereas for $U_0 \gg E_F$, the dice lattice ($\alpha=1$) has a larger utility. 
\end{abstract}

\maketitle

\section{Introduction}
After the remarkable discovery of being able to extract graphene monolayer \cite{Nov1,ADM}, the realization of a perfect two dimensional (2D) material could be achieved. On a more fundamental note, Dirac physics in realistic systems became one of the most explored topics in the field of condensed matter physics. Graphene, a two dimensional single layer of honeycomb lattice (HCL) formed of carbon (C) atoms, manifests a low-energy spectrum which is linear in momentum and obeys a (pseudospin) $s = 1/2$ Dirac-Weyl equation. The conduction band meets the valence band at six corner points of the hexagonal Brillouin zone (BZ), known as the Dirac points.
Due to the presence of the Dirac-like energy spectrum of quasiparticles, a number of fascinating physical phenomena, such as anomalous quantum Hall effect \cite{YZ,Nov2}, chiral tunneling \cite{MIK,CWJB}, Klein paradox \cite{MIK,MZ}  have emerged as distinctive properties of graphene. Furthermore, it  has generated tremendous
interest in different fields, such as electronics \cite{AHCN}, optoelectronics \cite{FB,PA}, and spintronics \cite{AGM,DP,BG} in the recent years. Moreover, it has been predicted that a graphene normal metal-superconductor (NS) and a normal metal-insulator-superconductor (NIS) junction can exhibit specular Andreev reflection in addition to the usual retro-reflection. The conductance characteristics of the junction systems and their utilities for heat transport are of prime importance in this work.

 An interesting variant of the honeycomb structure, or in other words, graphene, exists which goes by the name $\alpha-T_3$-lattice \cite{MIK2}, where the quasiparticles obey the Dirac-Weyl equation  with  pseudospin $s = 1$. This  $\alpha- T_3$ lattice is a special honeycomb like structure with an additional atom sitting at the corner of each hexagon. A unit cell of the $\alpha- T_3$ lattice comprises of three non-equivalent lattice sites, where two sites, generally known as the $rim$ sites, are located at the corner points of HCL. The other lattice site, called as the $hub$ site, is situated at the center of HCL.  The $rim$ sites are connected to the three nearest neighbors (NNs), while the hub site is connected to six NNs. The hopping strength between the $hub$ site and one of the $rim$ sites is proportional to a parameter $\alpha$, as shown in Fig.\ref{Fig1} (upper panel). The strength of $\alpha$ may be considered as a tunable parameter and at the extremities of the range [0:1] lie graphene and a dice lattice respectively \cite{BS,JV,SEK,MR,DFU,JDM,DB1,MV}. The low-energy excitations of the $\alpha-T_3$ lattice near the Dirac points consist of three branches. Two of them linearly disperse with momentum, generally known as the conic bands, while the third, being non-dispersive energy band is termed as a flat band. All the six band-touching points (the so called Dirac points) in the first BZ lie on the flat band. 
 
 A dice lattice can be realized by growing trilayers of cubic lattices (e.g., SrTiO3/SrIrO3/SrTiO3) in the (111) direction \cite{FW}. Further, in the context of cold atom, a suitable arrangement of three counterpropagating pairs of laser beams [19] can produce an optical dice lattice. Moreover, the Hamiltonian of a Hg$_{1-x}$Cd$_x$Te quantum well at a certain critical doping can also be mapped onto the $\alpha-T_3$ model in the intermediate regime (between dice and graphene), corresponding to a value of $\alpha =1/\sqrt{3}$ \cite{JD} where the band structure comprises of linearly dispersing conduction and valence bands, plus a flat band.

  A good number of physical quantities, such as the Berry-phase-dependent direct current (DC) Hall conductivity \cite{EI}, dynamical optical conductivity \cite{EI}, magneto-optical conductivity, the Hofstadter butterfly \cite{JN, TB, TK}, Berry-phase-modulated valley-polarized magnetoconductivity \cite{SK}, the photoinduced valley and electron-hole symmetry breaking \cite{BD} in the $\alpha-T_3$ lattice have been studied recently. Further, other properties, such as the conductivity \cite{TL, JW}, super-Klein tunneling \cite {RS, DF, EJ, YB}, gap generation and flat band catalysis \cite{EV, VP}, non-linear optical response \cite{LC}, topological phase transition\cite{DB, TKG, OP, SC, JF},  electronic and optical properties under radiation \cite{AI, LZ}, flat-band induced diverging DC conductivity \cite{MV}, and the thermoelectric performance of a nanoribbon \cite{MW} of $\alpha-T_3$ lattice have also been explored. Moreover, nontrivial topology\cite{ET, ZG, TN, ZL, MG, NS}, the diamagnetic\cite{Mc} (at $\alpha = 0$) to paramagnetic\cite{BS, JV} (at $\alpha = 1$) transition in the orbital magnetic responses of the lattice  have also been reported.

Recently, quantum transport studies through such junction systems have gained much more impetus in the field of  developing nano-scale devices \cite{SB}. The junction devices have compelling applications in the fields of thermoelectric, thermometric, solid-state cooling, etc. Previously, numerous studies have been performed \cite{NM,JM,MY1,MY2,PK,KP,PK1,PK2}, where the junction devices are found to be  useful in a wide range of applications in low-temperature thermometry and electronic cooling \cite{AVF,AMC,NAM}. The recent advancement in the field of thermoelectric physics in small-scale junction devices have provided new directions for manufacturing self-cooling devices, thermopower devices, etc.

Motivated by the above prospects of junction system at the nanoscale, we perform an extensive study of the electrical and the thermal transport in $\alpha-T_3$ based NIS junction system using the modified Blonder-Tinkham-Klapwijk (BTK) theory that adequately captures the low-energy transmission characteristics. In particular, we  compute the differential conductance of the $\alpha-T_3$ NIS junction as a function of the effective barrier potential (defined in the next section) for different values of the parameter $\alpha$. Further, we calculate the Seebeck coefficient, charge conductance, thermal conductance, figure of merit, maximum power, efficiency at maximum power, thermal current of such $\alpha-T_3$ based NIS junction, and explore an interplay between $\alpha$ (to interpolate between graphene and a dice lattice) and the effective barrier potential on their performance of these junction systems.

This paper is organized as follows. In Sec. \ref{Sec2}, we present the basic information regarding the junction (Sec. \ref{Sec2A}) and the definitions of different thermoelectric coefficients (Sec. \ref{Sec2B}, Sec. \ref{Sec2C}, and Sec. \ref{Sec2D}) that we shall be studying here. Section \ref{Sec3} includes details of the numerical results and their corresponding discussions. Finally, we conclude and summarize our main results in Sec. \ref{Sec4}.

\section{Model and Formalism}
\label{Sec2}
\begin{figure}
	\begin{center}
		\includegraphics[width=85mm,height=57mm]{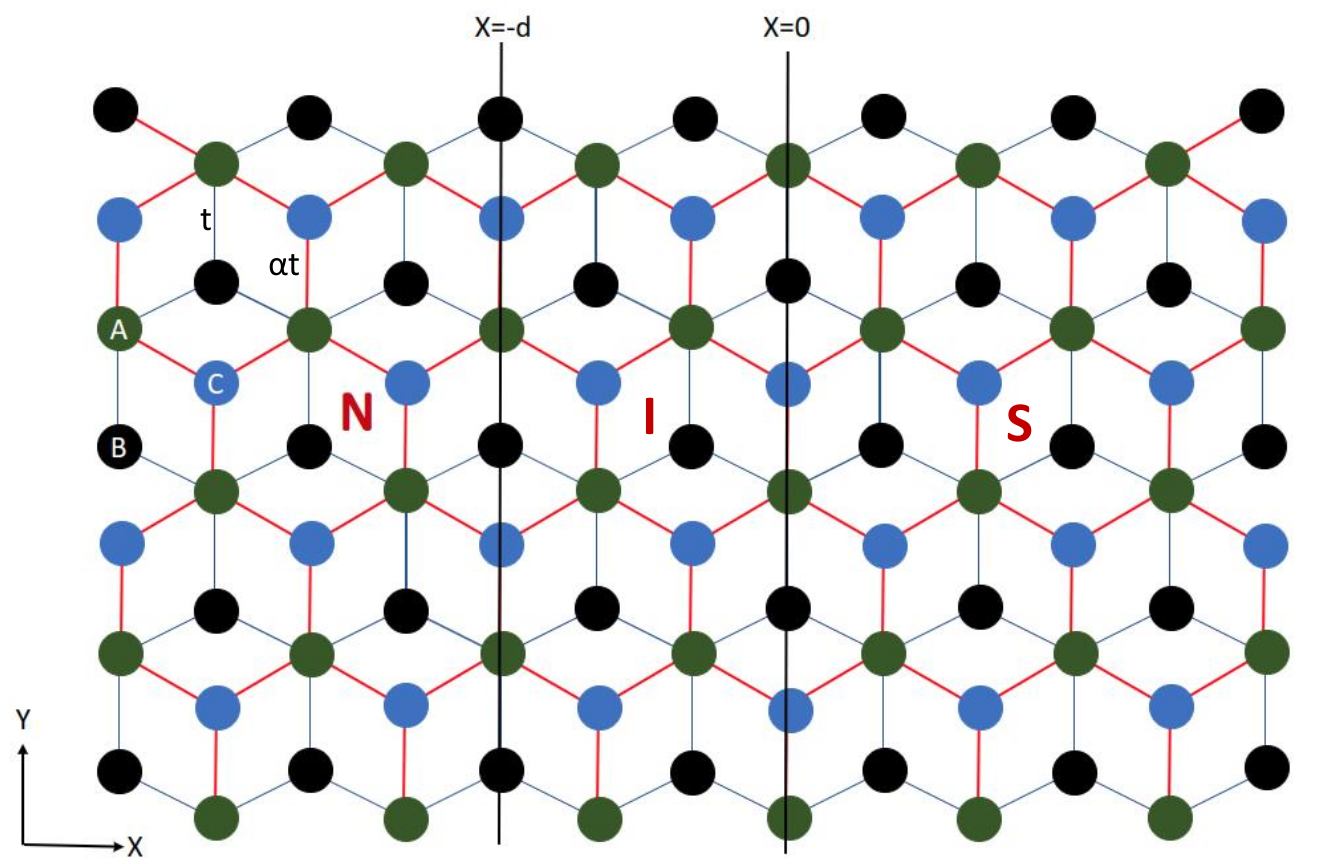}
		\includegraphics[width=85mm,height=57mm]{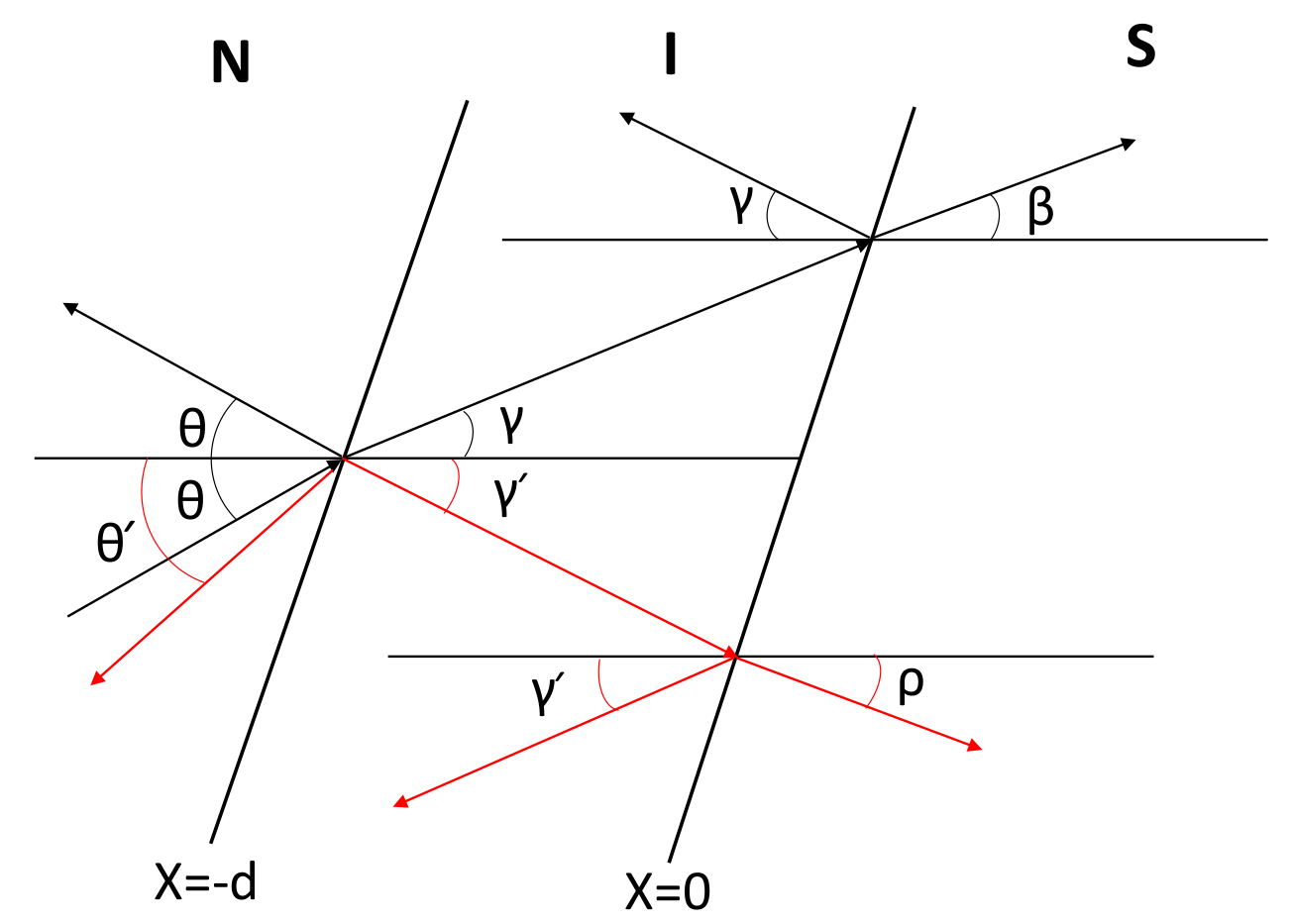}
		\caption {Upper panel: $\alpha-T_3$ lattice based normal metal insulator superconductor (NIS) junction. There are three atoms per unit cell:  sites A (green) and B (black), connected via hopping $t$, and an additional site C (blue) at the center of the hexagons, connected with A via a variable hopping parameter $\alpha t$. Lower panel: schematic illustration of the reflection and the transmission processes of the quasiparticles at the NIS junction. Black lines for electron like quasiparticles and red lines for hole like quasiparticles. }
		\label{Fig1}
	\end{center}
\end{figure}

This section initially presents all the essential information of our junction system, where the electron transport lies in the quasi-ballistic regime. In Fig.\ref{Fig1} (lower panel), we have shown the schematic illustration of all the reflection and the transmission processes of the quasiparticles at different regions of the NIS junction system. The later part includes the general description of thermoelectric coefficient, charge conductance, thermal conductance, figure of merit (FM), maximum power, efficiency at maximum power, and thermal current along with thermoelectric cooling for any junction system.
\subsection{Information of the junction system} 
\label{Sec2A} 
We consider an NIS junction made of  $\alpha -T_3$ sheet, occupying the $xy$ plane (see Fig.\ref{Fig1}(upper panel)). The left lead, i.e., the normal region (N) extends from $x = -\infty$ to $x = -d$ for all $y$. The middle region, i.e., the insulating regime ($I$) is characterized by a barrier potential $V_0$, extending from $x = -d$ to $x = 0$. The right superconducting lead (S) occupies the region $x \geq 0$. The local barrier in the insulating regime can be implemented either using by an external electric field or a local chemical doping \cite{No, Zh, KS, MI}. The superconductivity in the right lead can be induced via a proximity effect \cite{CW, AF}. It is to be noted  that the barrier region has sharp edges on both sides, thus the insulating regime has to satisfy the condition: $d \ll 2\pi/k_F$, where $k_F$ is the Fermi wave vector for an $\alpha -T_3$ lattice. The NIS junction of our consideration can be described by the Dirac-Bogoliubov$-$de Gennes (DBdG) equation \cite{CW, KS1, Zho}
\begin{equation}
	\label{eq1}
\begin{pmatrix}
H-E_F & \Delta_0 e^{i\varphi}\sigma_0\\
\Delta_0 e^{-i\varphi}\sigma_0 & E_F-\mathcal{T}H\mathcal{T}^{-1}
\end{pmatrix}
\begin{pmatrix}
u_e\\
v_h
\end{pmatrix}=\epsilon
\begin{pmatrix}
u_e\\
v_h
\end{pmatrix},
\end{equation}
where $E_F$ is the Fermi energy of system, $\varphi$ denotes the macroscopic phase in the superconducting regime, $\mathcal{T} = \sigma_x \otimes \tau \textit{C}$ is the time-reversal operator with $\sigma_x$ being the Pauli matrix, $\textit{C}$ being the operator of complex conjugation and $\tau = \begin{pmatrix}
1 & 0 & 0\\
0 & -1 & 0\\
0 & 0 & 1
\end{pmatrix}$.  Further, $\epsilon$ denotes the excited energy of electron and hole, $u_e$ and $v_h$ are the electron (electron-like) and hole (hole-like) wave functions in
the normal (superconducting) regime, respectively and $\sigma_0$ denotes a unit matrix. The
zero temperature superconducting gap, $\Delta_0$ is induced in the $\alpha-T_3$ lattice by addressing a conventional $s$-wave superconductor.  The Hamiltonian in
 $\alpha-T_3$ lattice can be written as,
\begin{equation}
H= \begin{pmatrix}
H_+ & 0\\
0 & H_-
\end{pmatrix},
\end{equation}
where $H_\pm = \hbar v_F \textbf{S}.\textbf{k} + U(x)$ with,
\begin{equation}
S_x= \pm
\begin{pmatrix}
0 & \cos\phi & 0\\
\cos\phi & 0 & \sin\phi\\
0 & \sin\phi & 0
\end{pmatrix}
\end{equation} and
\begin{equation}
S_y= -i
\begin{pmatrix}
0 & \cos\phi & 0\\
-\cos\phi & 0 & \sin\phi\\
0 & -\sin\phi & 0
\end{pmatrix}.
\end{equation}
Here, $v_F$ is the Fermi velocity, the label $\pm$ denote the $K$ and $K'$ valleys, respectively. The angle $\phi$ is related to the strength of coupling parameter $\alpha$ via $\alpha = \tan\phi$. The potential $U(x)$ gives rise to the relative shift in Fermi energies for normal, insulating, and superconducting regimes of  $\alpha -T_3$ sheet, where $U(x)$ is modeled as $U(x) = -U_0\Theta(x) + V_0 \Theta(-x) \Theta(x+d)$ with $\Theta$ being the Heaviside step function. Now, we introduce a dimensionless barrier strength via \cite{KS1},
\begin{equation}
\chi = V_0d/\hbar v_F,
\end{equation}
which is going to play a pivotal role in all our subsequent discussions. We consider a thin barrier, such that with $V_0 \to \infty$
and $d \to 0$, $\chi$ remains finite. For a NS (normal-superconductor) junction $\chi$ vanishes. The gate potential $U_0$ can mimic the Fermi surface mismatch between the normal and superconducting regimes. It is to be noted that the mean-field conditions for superconducting regime are satisfied as
long as $\Delta_0 \ll (U_0 +E_F)$. Thus, in principle, for large $U_0$
one can have regimes where $\Delta_0 \geq E_F$.

\begin{figure}
\includegraphics[width=0.4\textwidth]{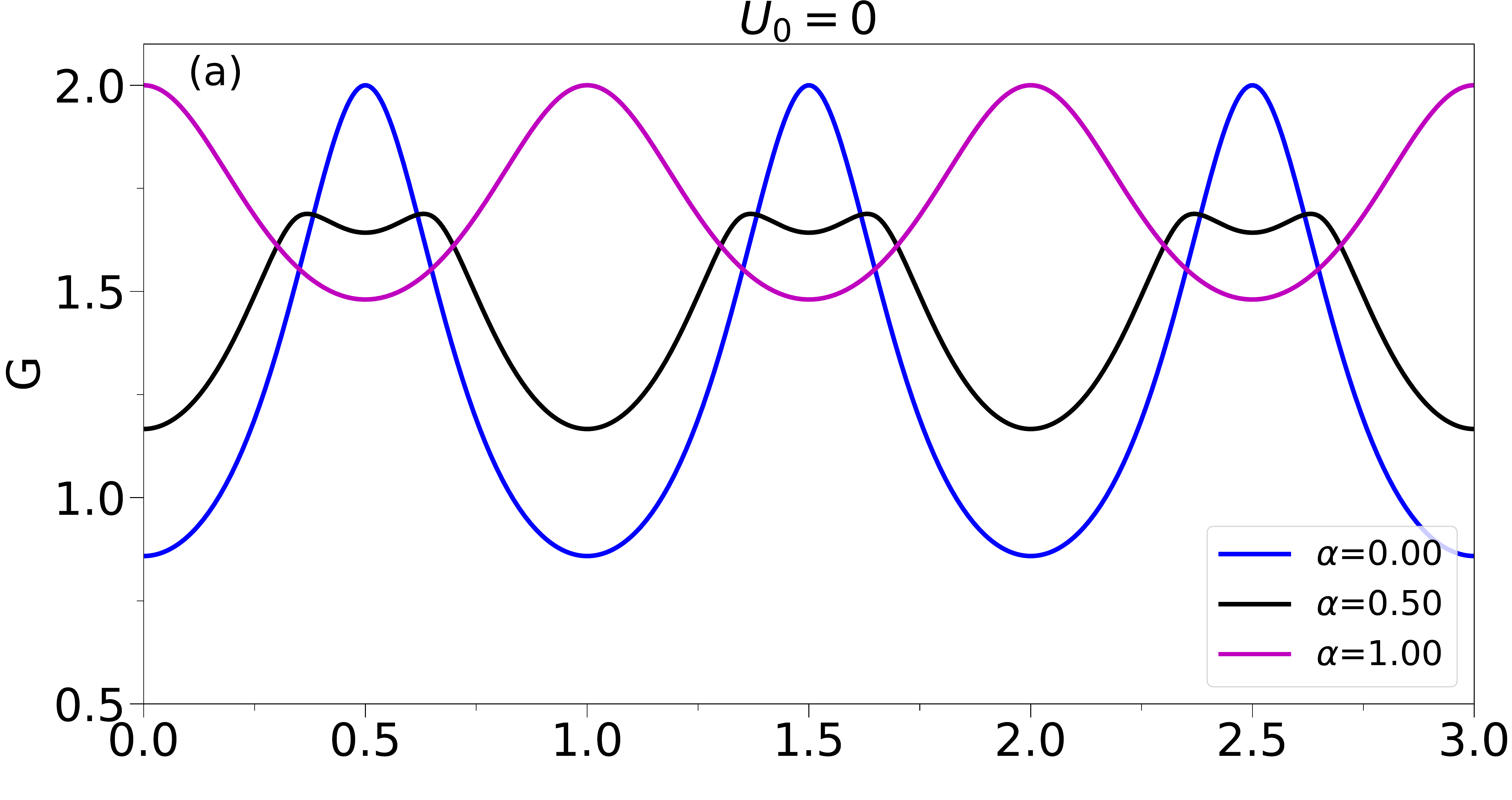}
\includegraphics[width=0.4\textwidth]{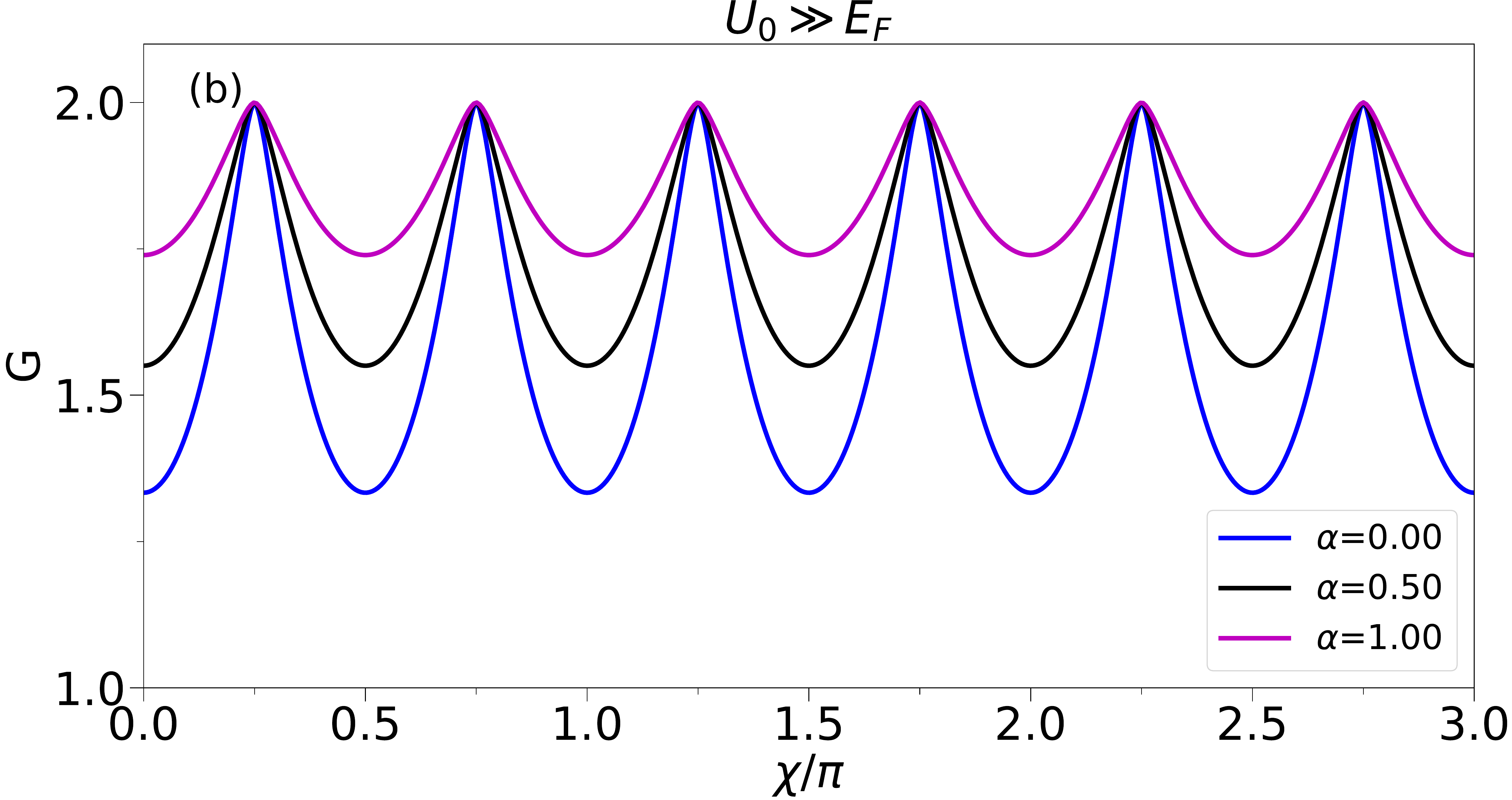}
\caption{Differential conductance $G$ (in unit of $G_0$) as a function of the effective barrier potential $\chi$ (a) for $U_0$ = 0 and $E_F \gg \Delta_0$. (b) for $U_0 \gg E_F$ and $E_F \gg \Delta_0$.}
\label{Fig2}
\end{figure}

Because of the time-reversal symmetry of the $\alpha-T_3$ lattice ($\mathcal{T}H\mathcal{T}^{-1}= H$), Eq.(\ref{eq1}) can be decoupled into two sets of six equations with the  form,
\begin{equation}
	\label{eq6}
\begin{pmatrix}
H_\pm-E_F & \Delta_0 e^{i\rho}\sigma_0\\
\Delta_0 e^{-i\rho}\sigma_0 & E_F-H_\pm
\end{pmatrix}
\begin{pmatrix}
u_e\\
v_h
\end{pmatrix}=E
\begin{pmatrix}
u_e\\
v_h
\end{pmatrix}.
\end{equation}
We consider only $H_+$ because of the valley degeneracy. The energy dispersion can be written as $E = \sqrt{(E_F+U(x)\pm \hbar v_F |k|)^2+\Delta_0^2\Theta(x)}$, with $|k|=\sqrt{k_x^2+k_y^2}$ in both the normal and superconducting regions. For a given $E$ and $k_y$, the four eigenstates in the normal region ($ -\infty < x \leq -d$, $\Delta_0=0$) are obtained as  
\begin{equation}
\psi_e^\pm = 
\begin{pmatrix}
  \pm e^{\mp i\theta}\cos \phi\\
   1\\
   \pm e^{\pm i \theta}\sin\phi\\
   0\\
   0\\
   0
\end{pmatrix}e^{i(\pm k_x^ex + k_yy)},
\end{equation}
and
\begin{equation}
\psi_h^\pm =
\begin{pmatrix}
 0\\
 0\\
 0\\
 \mp e^{\mp i\theta'}\cos \phi\\
 1\\
 \mp e^{\pm i \theta'}\sin\phi
\end{pmatrix} e^{i(\pm k_x^hx + k_yy)}.
\end{equation}
The state $\psi_e^+$ ($\psi_h^+$) denotes the electron (hole) moving along $+x$ direction (i.e., towards the normal metal-insulator (NI) junction), while $\psi_e^-$ ($\psi_h^-$) denotes the electron (hole) moving
along $-x$ direction (away from the NI junction). Further, the angles
$\theta = sin^{-1}[\hbar v_F k_y /(E + E_F)]$ and $\theta'= sin^{-1}[\hbar v_F k_y /(E - E_F)]$ denote the incident angle of an electron and the reflected angle of the corresponding hole, respectively. The wave vector $k_x^e$ ($k_x^h$) $= \frac{E+(-)E_F}{\hbar v_F}\cos\theta$ $(\theta')$ represents  the longitudinal wave vector of the electron (hole). 

\begin{figure}
\includegraphics[width=0.47\textwidth]{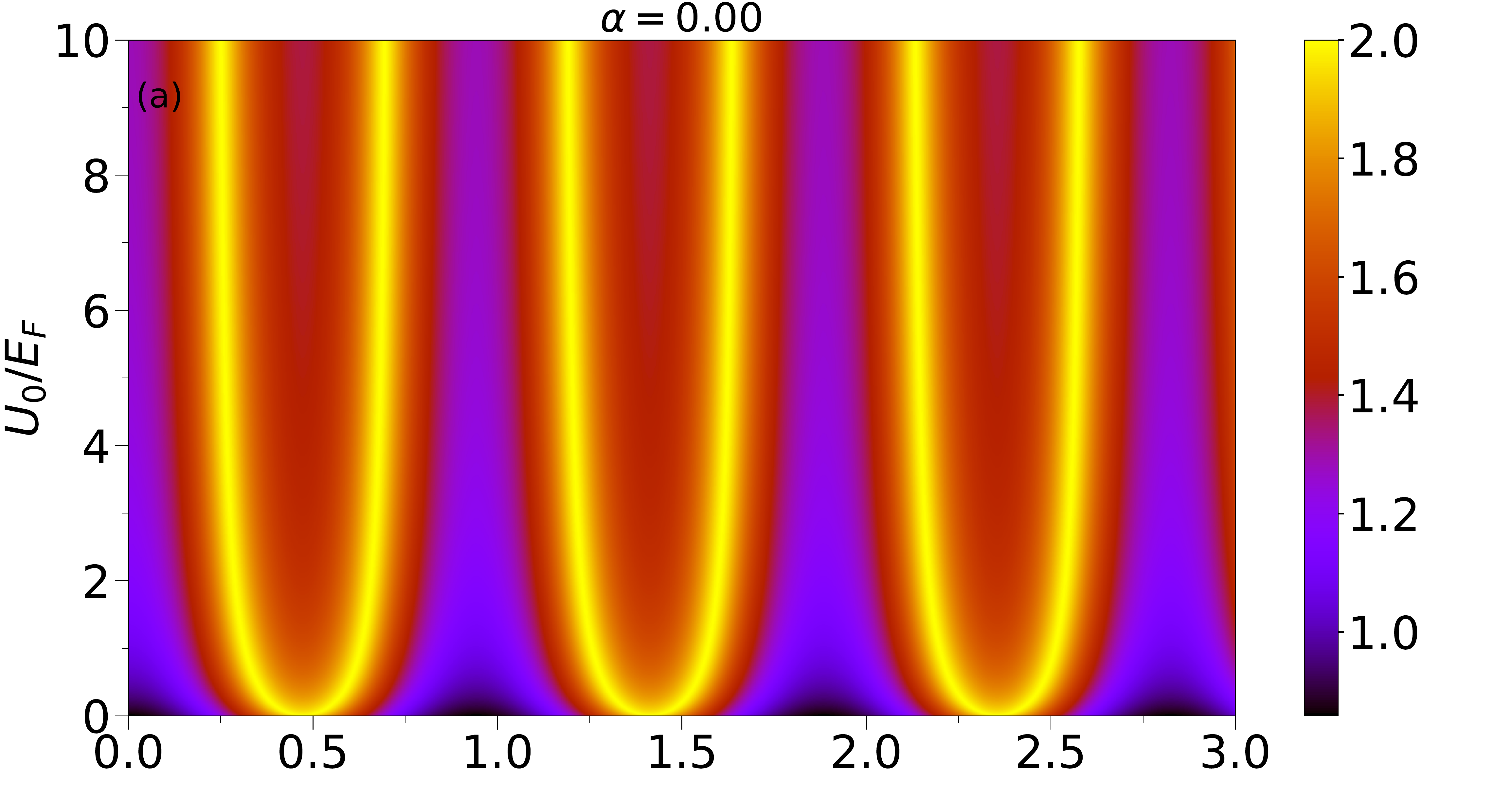}
\includegraphics[width=0.47\textwidth]{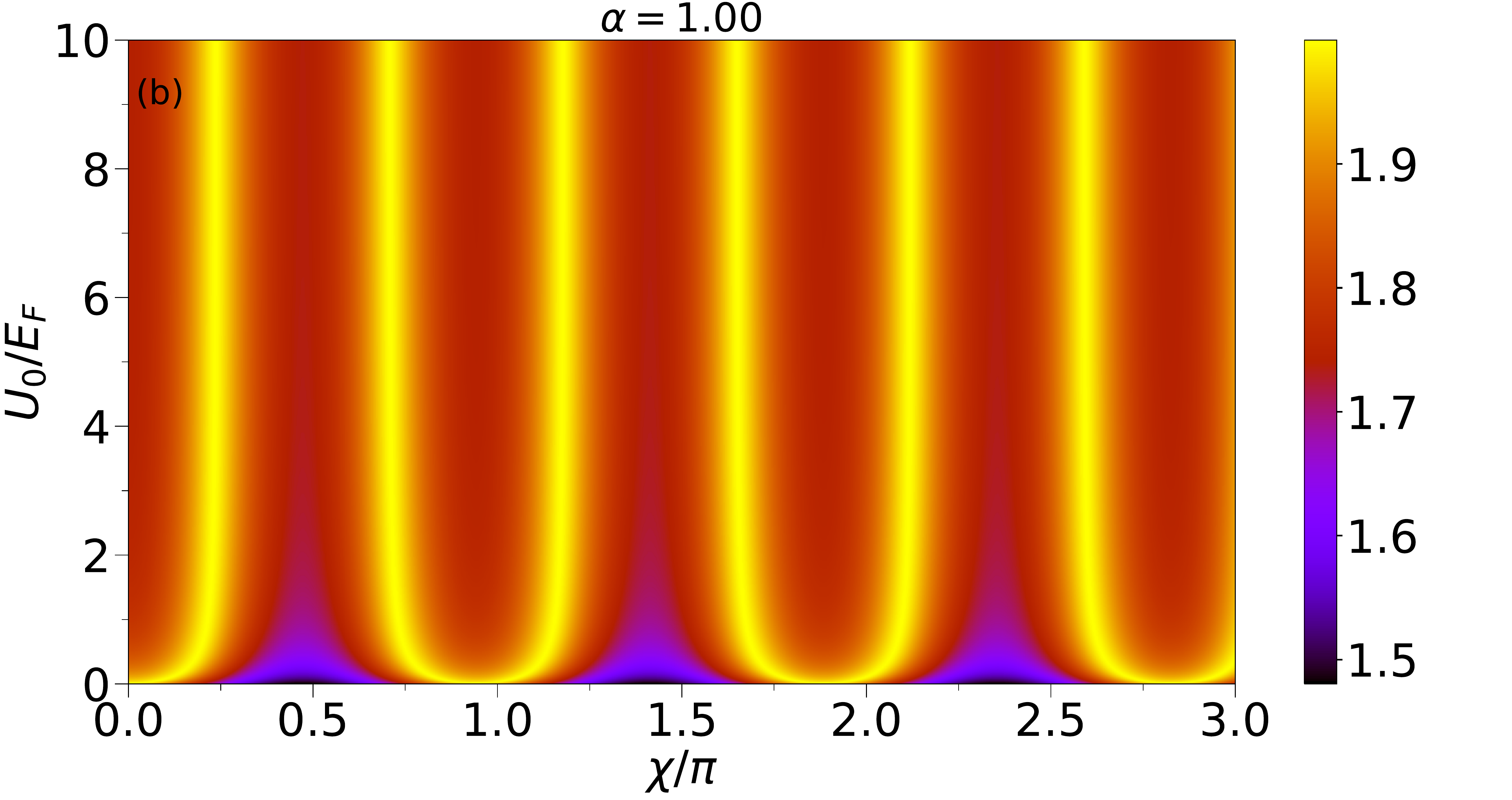}
\caption{The variation of differential conductance $G$ (in unit of $G_0$) as a function of the effective barrier potential $\chi$ and gate voltage in the superconducting region $U_0$ (a) for $\alpha = 0$, (b) for $\alpha=1$.}
\label{Fig3}
\end{figure}

The wave functions in the superconducting region ($x\geq0$) are obtained as,
\begin{equation}
\psi_S^\pm=
\begin{pmatrix}
e^{\pm i \beta} e^{\pm 2i \partial}\\
\pm \frac{1}{\cos\phi}e^{\pm i \beta} e^{\pm i \partial}\\
\tan\phi e^{\pm\beta}\\
e^{-i\rho}e^{\pm 2i \partial}\\
\pm \frac{1}{\cos\phi}e^{-i\rho}e^{\pm i \partial}\\
\tan\phi e^{-i\rho}
\end{pmatrix} e^{(\pm ik_0x + ik_yy - \kappa x)},
\end{equation}
with $k_0 = U_0/ \hbar v_F$, $\kappa = (\Delta_0/\hbar v_F)/\sin\beta$ and 
$\beta=  \cos^{-1} (\frac{E}{\Delta_0})$ for $E<\Delta_0$,
   $\beta= -i\cosh^{-1} (\frac{E}{\Delta_0})$ for $E >\Delta_0$.
 
Similarly, in the intervening region ($-d< x < 0$), one can obtain,
\begin{equation}
\psi_I^{e\pm} = 
\begin{pmatrix}
\pm e^{\pm i\gamma} \cos\phi\\
1\\
\pm e^{\mp i\gamma} \sin\phi\\
0\\
0\\
0
\end{pmatrix} e^{(\pm i k_{x_I}x + i k_{y}y)},
\end{equation}
 and
\begin{equation}
\psi_I^{h\pm} = 
\begin{pmatrix}
0\\
0\\
0\\
\mp e^{\mp i\gamma'} \cos\phi\\
1\\
\mp e^{\pm i\gamma'} \sin\phi
\end{pmatrix} e^{(\pm i k_{x_I}'x + i k_{y}y)},
\end{equation}
with $k_{x_I} = \frac{[E+(E_F-V_0)]\cos\gamma}{\hbar v_F}$ and $k_{x_I}' = \frac{[E-(E_F-V_0)]\cos\gamma'}{\hbar v_F}$ being the corresponding wave vectors. 
The eigenstate $\psi_I^{e+}$ ($\psi_I^{h+}$) denotes the electron (hole) moving along $+x$ direction (i.e., towards the insulator-superconductor (IS) junction), where $\psi_I^{e-}$ ($\psi_I^{h-}$) denotes the electron (hole) moving along the $-x$ direction (i.e., away from the IS junction).
Thus, the wave functions in the normal, insulating and superconducting regimes can be written as,
\begin{eqnarray}
\psi_N &=& \psi_N^{e+} + r\psi_N^{e-}+r_A\psi_N^{h-},\\\nonumber
\psi_I &=& p\psi_I^{e+} + q\psi_I^{e-}+m\psi_I^{h+}+n\psi_N^{h-}\\\nonumber
\psi_S &=& t\psi_S^+ + t'\psi_S^-,
\end{eqnarray}
where $r$ and $r_A$ are the amplitudes of normal and Andreev reflections, respectively, in the normal regime; $p$ and $q$ are the amplitudes of the incoming and reflected electrons in the insulating regime, $m$ and $n$  denote the amplitudes of incoming and reflected holes in the insulating region. Further, $t$ and $t'$ correspond to coefficients of transmission to the superconducting regime as electron-like and hole-like quasiparticles, respectively.

 The wave functions must satisfy the following boundary conditions: 
\begin{equation}
\begin{split}
\psi_N|_{x=-d} = \psi_I|_{x=-d},\\
\psi_I|_{x=0} = \psi_S|_{x=0}.
\end{split}
\end{equation}
Using the boundary conditions, one can obtain the normal reflection coefficient $r$ and Andreev reflection coefficient $r_A$. With some straightforward, although cumbersome algebra, we find that in the limit of a thin barrier, the expressions for $r$ and $r_A$ depend on the dimensionless barrier strength $\chi$ as,
\begin{equation}
r = \frac{(X_5L_6 -1)}{(1-L_6X_6)}e^{-2i\chi},
\end{equation}
and
\begin{equation}
r_A = m \frac{e^{-i\chi}}{e^{ik_x^hd}}+n \frac{e^{i\chi}}{e^{ik_x^hd}},
\end{equation}
where, 
\begin{eqnarray}
L_6=\frac{L_5 + e^{2ik_{xI}d}}{L_5X_1 + X_2e^{2ik_{xI}d}} & \hspace{1mm};\hspace{1mm}&
L_5=\frac{X_2L_4-1}{1-X_1L_4},\\\nonumber
L_4=\frac{L_3e^{2i\beta}-1}{L_3e^{2i\beta}+1}\hspace{1mm}&;\hspace{1mm}&
L_3=\frac{1+L_2}{1-L_2},\\\nonumber
L_2=\frac{L_1+1}{L_1X_3+X_4}\hspace{1mm}&;\hspace{1mm}&
L_1=\frac{X_7-X_8}{X_9-X_7}e^{2i\chi},
\end{eqnarray}
and
\begin{eqnarray}
X_1/X_2 & =& \pm e^{\mp i\gamma}\cos^2\phi \pm e^{\pm i\gamma}\sin^2\phi,\\\nonumber
X_3/X_4 &=& \mp e^{\pm i\gamma'}\cos^2\phi \mp e^{\mp i\gamma'}\sin^2\phi,\\\nonumber
X_5/X_6 &=&\pm e^{\mp i\theta}\cos^2\phi \pm e^{\pm i\theta}\sin^2\phi,\\\nonumber
X_7 &=& e^{i\theta'}\cos^2\phi + e^{-i\theta'}\sin^2\phi,\\\nonumber
X_8/X_9& =& \pm e^{ \mp i\gamma'}\cos^2\phi \pm e^{\mp i\gamma'}\sin^2\phi.\\
\end{eqnarray}

 Using the BTK formalism, the differential conductance at zero temperature is given by,
\begin{equation}
	\label{eq19}
	G = G_0 \int_0^{\frac{\pi}{2}}\tau_1(E,\theta)\cos\theta d\theta,
\end{equation}
with $\tau_1(E,\theta)=(1-R+R_A)$, $R=r^*r$, and $R_A=r_A^*r_A$.
Considering the two-fold spin and valley degeneracies, $G_0 = \frac{4e^2}{h}N(E)$ is the ballistic conductance with $N(E) = \frac{W(E + E_F)}{\pi \hbar v_F}$ being the transverse modes in the $\alpha-T_3$ lattice with a total width, $W$.

\begin{figure}
\includegraphics[width=0.45\textwidth]{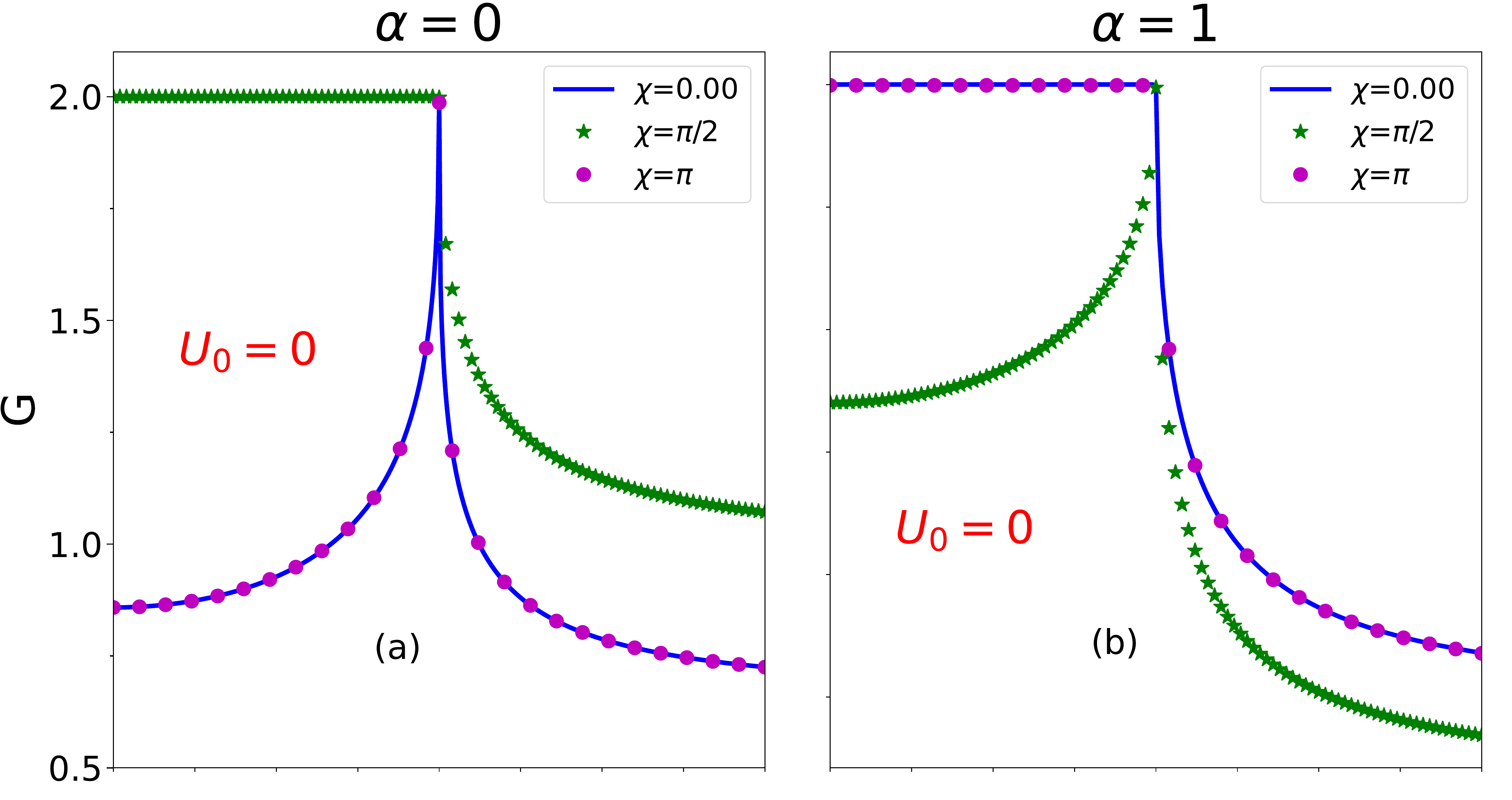}
\includegraphics[width=0.45\textwidth]{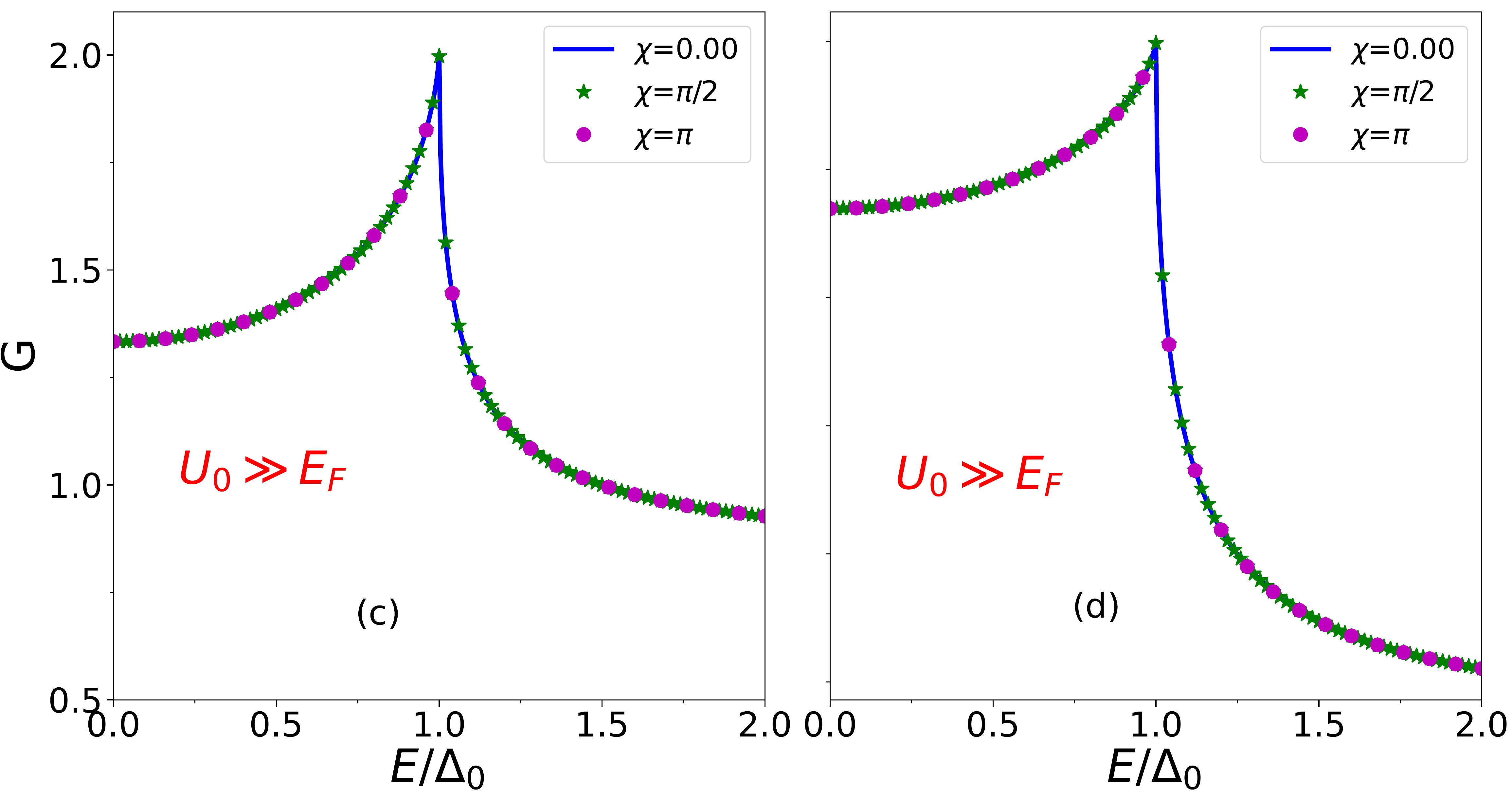}
\caption{Differential conductance $G$ (in unit of $G_0$) as a function of the bias voltage $E/\Delta_0$ for different values of $\chi$, for (a) $\alpha =0$, $U_0=0$, (b) $\alpha =1$, $U_0=0$, (c) $\alpha =0$, $U_0 \gg E_F$, and (d) $\alpha =1$, $U_0 \gg E_F$.}
\label{Fig4}
\end{figure}

\subsection{Seebeck coefficient}
\label{Sec2B}
Here, we present the theory to calculate the Seebeck coefficient. A temperature difference
between two dissimilar materials produces a voltage difference, and this phenomenon is known as the Seebeck effect.
For our system,  the normal and the superconducting leads are subjected to a temperature difference $\delta T$ and there will be a voltage difference $\delta V$ (say) between the two leads. The thermopower or the Seebeck coefficient $S$ is defined as the voltage induced per unit temperature difference in an open circuit condition. Hence we can write
\begin{equation}
	S = (\frac{\delta V}{\delta T})_{I=0}.
\end{equation}
We consider the left and right electrodes as independent temperature
reservoirs, where the left and right electrodes are kept at temperature $T^N = T - \delta T/2$ and $T^S = T + \delta T/2$, respectively. The populations of electrons in the left and right leads are described by the Fermi-Dirac distribution functions, namely, $f^N$ and $f^S$, respectively, where $E_F^N = E_F^S$ at zero external bias.

In an open circuit condition, let us now consider an extra infinitesimal current induced by an additional
voltage $\delta V$ and the temperature difference $\delta T$ across the junction. The currents induced by $\delta T$ and $\delta V$ are given by, $(dI)_T = I ( E_F^N , T^N , E_F^S =E_F^N , T^S = T^N + \delta T )$ and $( dI )_V = I ( E_F^N , T^N , E_F^S = E_F^N + e\delta V, T^S = T^N )$. Suppose the induced current cannot flow in an open circuit condition, where $( dI )_T$ should counter-balances $( dI )_V$. Hence, we can write,
\begin{equation}
	dI = (dI)_T + (dI)_V = 0.
\end{equation}
Performing the first order expansion of Fermi-Dirac distribution function in $( dI )_ T$ and $( dI )_V$ with energy shifted by Fermi energy, one can obtain the expression of Seebeck coefficient,
\begin{equation}
	S = \frac{\delta V}{\delta T}=\frac{\int \int dE d\theta \cos\theta E(E+E_F^N)\tau_1(E,\theta)\frac{\partial f}{\partial E}}{\int \int dE d\theta \cos\theta (E+E_F^N)\tau_1(E,\theta)\frac{\partial f}{\partial E}},
\end{equation}
where $\tau_1(E,\theta) = (1-R+R_A)$, and $f$ is the Fermi-Dirac distribution function.

\subsection{Maximum power, efficiency at maximum power, and figure of merit}
\label{Sec2C}
The most challenging part to fabricate a thermoelectric device is to find the optimal conditions which ensure the operation of the device with maximum power output at the best possible efficiency. We are also looking for a condition to derive maximum power output from our NIS junction system, where the maximum power is given by, 
\begin{equation}
\label{eq23}
P_{max} = \frac{1}{2}S^2 \mathcal{L} (\Delta T)^2
\end{equation}
with $\mathcal{L} = \int \int \tau_1(E, \theta) (E+E_F)(-\frac{df}{dE}) dE \cos\theta d\theta $ \cite{AM}. Now, for optimal operation of the thermoelectric device, its function with the highest possible efficiency is also have to be considered. The efficiency ($\eta$) of the thermoelectric system is taken to be the ratio of the power to the
thermal current or in other words the efficiency of a device is defined as the ratio of useful output power to the input power. Since our main focus is to look into the operation of the NIS junction system at the maximum power, we, therefore, concentrate upon the efficiency calculated at the maximum power which is given
by \cite{AM,No1}, 
\begin{equation}
\label{eq24}
\eta (P_{max}) =\frac{\eta_c}{2} \frac{zT}{(2+zT)}
\end{equation}
where $\eta_c$ is the Carnot efficiency.
The efficiency of the system depends upon a quantity
called the figure of merit ZT, which is defined as,
\begin{equation}
\label{eq25}
	ZT = \frac{S^2G_c}{G_{Th}}T,
\end{equation}
where $S$ is the Seebeck coefficient, $G_c$ is the charge conductance, $G_{Th}$ is the thermal conductance, and $T$ is the absolute temperature. $G_c$ can be calculated from the relation
\begin{equation}
	G_c= \frac{1}{2eR_NE_F^N}\int\int \tau_1(E,\theta) \Bigl(-\frac{\partial f}{\partial E}\Bigl)(E+E_F^N)dE \cos\theta d\theta,
\end{equation}
where $R_N$ is the normal state resistance, defined as $R_N = \frac{1}{2e^2N_0v_F^NA_r}$ (2 is the spin degeneracy factor, $N_0$ is the density of states at the Fermi level, $V_F^N$ is the Fermi velocity, and $A_r$ is the area of contact). The thermal conductance, $G_{Th}$ can be calculated from the
relationship $G_{Th} = \frac{dJ_{NS}}{dT}$, where $J_{NS}$ is the thermal
current flowing from the normal lead to the superconducting lead. In the next subsection, we present how the thermal current and the thermal conductance can be calculated.

\subsection{Thermoelectric cooling and thermal conductance}
\label{Sec2D}
 The flow of electrons can also transport thermal energy through the junction, which is responsible for the thermal current. The thermal current is defined as the rate at which the thermal energy flows from the left lead to the right lead, where the left electrode serves as the cold reservoir and the right one serves as the hot reservoir. An external bias voltage $V_B = (E_F^N - E_F^S)/e$  drives the electrons to flow from  normal to superconducting lead. Thus, the electron removes the heat energy from the normal lead and transfers it to the superconducting lead which further
makes the cold reservoir (normal) cool. The energy conservation allows us to write,
\begin{equation}
	\begin{split}
		J_{NS}(E_F^N,T^N;E_F^S,T^S)+I_{NS}(E_F^N,T^N;E_F^S,T^S)V_B=\\
		J_{SN}(E_F^N,T^N;E_F^S,T^S).
	\end{split}
\end{equation}
Using the analogy between electronic charge current and electronic thermal current, one can write the outbound energy flow rate from normal lead to  superconducting lead as,
\begin{equation}
	\begin{split}
		J_{NS}=\frac{1}{2e^2R_NE_F^N}\int\int(E-eV_B)(E+E_F^N)\tau'(E,\theta)\\
		[f^N(E-eV_B,T^N)-f^S(E,T^S)]dE\cos\theta d\theta.
	\end{split}
\end{equation}
Similarly, the reverse, that is the rate at which the superconducting lead receives the thermal energy, is written as
\begin{equation}
	\begin{split}
		J_{SN}=\frac{1}{2e^2R_NE_F^N}\int\int E(E+E_F^N)\tau'(E,\theta)\\
		[f^N(E-eV_B,T^N)-f^S(E,T^S)]dE\cos\theta d\theta,
	\end{split}
\end{equation}
where the energies are shifted by Fermi energy and 
$\tau'(E,\theta) = (1-R-R_A)$.

The thermal conductance $G_{Th}$ can be obtained from the temperature derivative of $J_{NS}$, that is, $\frac{dJ_{NS}}{dT}$. This junction system can be regarded as the electronic cooling device only when $J_{NS} > 0$, which indicates that it is competent to remove the heat from cold reservoir and hence making it cooler. The thermal conductance, $G_{Th}$ is given by the form,
\begin{equation}
	\begin{split}
		G_{Th}=\frac{1}{2e^2R_NE_F^N}\int\int E(E+E_F^N)\tau'(E,\theta)\\
		\Bigl[\frac{f^N(E-eV_B,T^N)}{dT}-\frac{f^S(E,T^S)}{dT}\Bigl]dE\cos\theta d\theta
	\end{split}
\end{equation}

\begin{figure}
\centering
\includegraphics[width=0.47\textwidth]{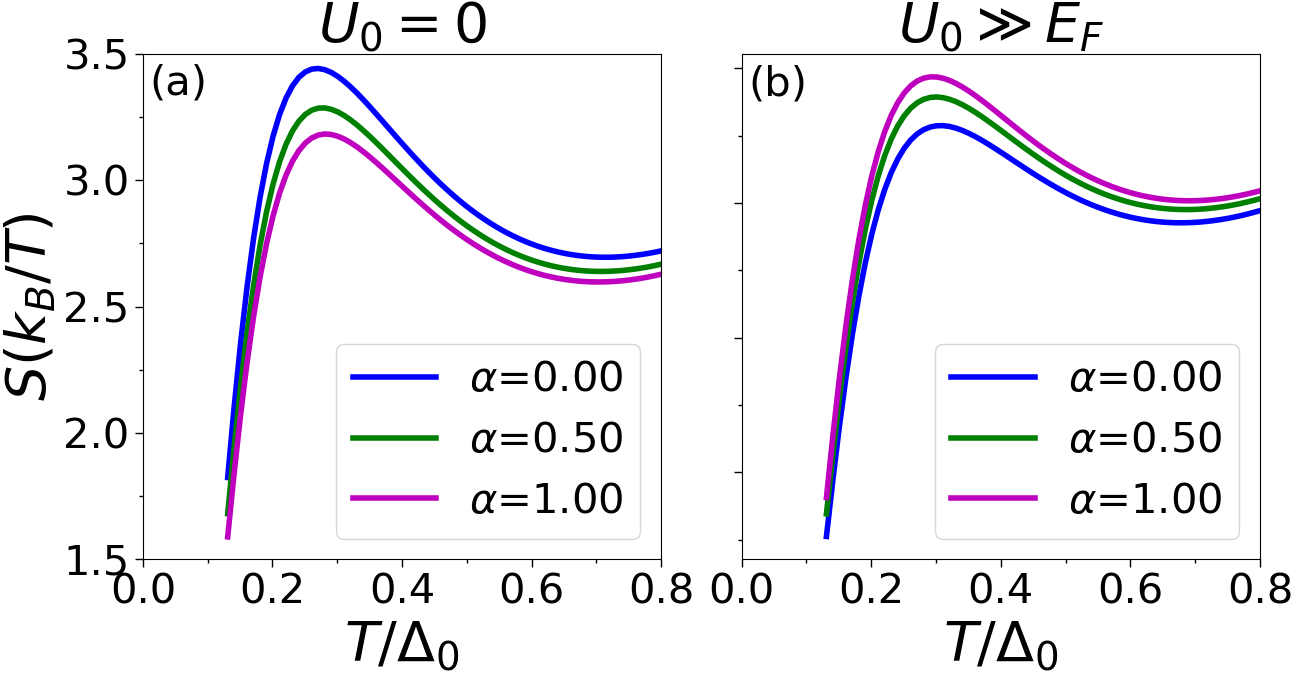}
\caption{The variation of Seebeck coefficient, $S (k_B/e)$ as a function of temperature (scaled by $\Delta_0$), for different values of $\alpha$ (a) for $U_0 = 0$ and (b) for $U_0 \gg E_F$, for a fixed value of $\chi$, where $G$ is maximum.}
\label{Fig5}
\end{figure}

\section{Results and Discussions}
\label{Sec3}
Here, we present numerical results of zero temperature differential conductance and different thermoelectric properties of the $\alpha-T_3$ based NIS junction.  Putting
things in perspective, we have considered some reasonable values of $\Delta_0$ at 1meV, and $E_F =50\Delta_0$ (in the subsequent discussions, Fermi energy of the normal metal regime will consider as $E_F$). For thermoelectric studies, the temperature of the cold lead is kept at
$T$,  whereas the hot lead is kept at $T + dT$ with $dT \ll T$. We have varied the temperature in such a range so that the superconductivity is not destroyed.
The dimension of the system is in the range of a few nanometer.
\begin{figure}
\includegraphics[width=0.45\textwidth]{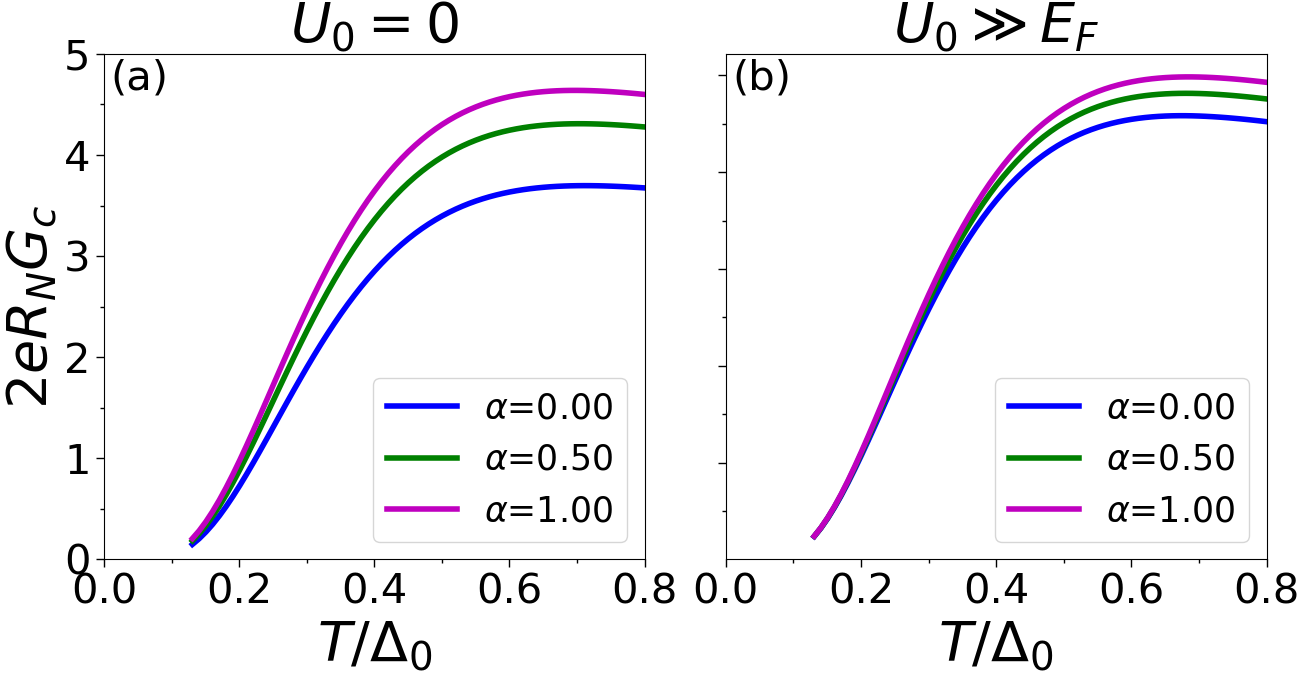}
\caption{The variation of the dimensionless charge conductance, $2eR_NG_c$ as a function of temperature (scaled by $\Delta_0$), for different values of $\alpha$ (a) for $U_0 = 0$ and (b) for $U_0 \gg E_F$.}
\label{Fig6}
\end{figure}
\\

\textbf{\textit{A. Differential Conductance (at T = 0 K)}:}
The differential conductance $G$ (at zero temperature)  as a function of  effective barrier potential $\chi$ for different values of $\alpha$ is shown in Fig.\ref{Fig2}. We consider two cases: $(i)$ $U_0=0$, $E_F \gg \Delta_0$, the Fermi energy of the superconducting region is taken as, $E_F^S = E_F^N$, and $(ii)$ $U_0 \gg E_F, E_F \gg \Delta_0$ i.e. the Fermi energy of superconducting region is much greater than the Fermi energy of the normal region. We have chosen $U_0 \gg E_F$ so that the leakage of Cooper pairs from the S to N regions can be safely neglected \cite{CW,HLi,Ouy}.

Fig.\ref{Fig2} shows Fabry-Perot like oscillation in the conductance pattern for both the cases indicating electron interferences. For case $(i)$, where the Fermi surfaces of the normal metal and the superconductor are aligned ($U_0 = 0$) and $E_F \gg \Delta_0$,  the oscillations are exactly in opposite phase for $\alpha =0$ and $\alpha = 1$. For $\alpha = 0$, the $\pi$ periodic oscillations show maxima at $\chi = (n+1/2)\pi$, where for $\alpha =1$, the $\pi$ periodic oscillations show  maxima at $\chi = n\pi$ (see Fig.\ref{Fig2}a) $(n=0,1,2,...)$. For a intermediate case, i.e., for any values of $\alpha$ lying between $0$ and $1$, a hump-like feature is obtained in the oscillations. As we know, for a Fabry-Perot interferometer the maxima and the minima depend on the path difference of the light passing through it, similarly here also, as we go from $\alpha =0$ (graphene) to $\alpha = 1$ (dice lattice), the electrons travel an extra path which comes out with exactly opposite in phases. Further, it is noticeable that the amplitudes of oscillations are different for different values of $\alpha$. As we know, for a Fabry-Perot oscillator the amplitudes depend upon the reflectivities of the first (analogous to NI interface of our system) and second mirror (analogous to IS interface of our system) of the resonator. 
Similarly, for the $\alpha-T_3$ lattice, the amplitudes of oscillation  decrease as we go from $\alpha =0$ to $\alpha =1$, indicating an increase in normal reflection coefficient ($r$), while a decrease in the Andreev reflection coefficient ($r_A$).
Now, we consider case $(ii)$, where we increase the value of $U_0$. The oscillations tend to occur in the same phase for all values of $\alpha$, as shown in Fig.\ref{Fig2}b and the maxima occur at $\chi=(n+1/2)\pi/2$ $(n=0,1,2,...)$. Here, the periodicity decreases to $\pi/2$ for all values of $\alpha$. For $U_0 \gg E_F$, the oscillations are exactly in the same phase, but their amplitudes are different. 

Fig.\ref{Fig3} shows a color plot of the differential conductance, ($G$) as a function of the effective barrier potential $\chi$ and the gate voltage in the superconducting region $U_0$. It is clear that for low values of $U_0$, the maxima of the conductance appear at different values of $\chi$ for both $\alpha=0$ (Fig.\ref{Fig3}a) and $\alpha=1$ (Fig.\ref{Fig3}b) cases. But for greater values of $U_0$, the maxima appears at the same value of $\chi$.

Further, in Fig.\ref{Fig4}, we show the tunneling conductance $G$ as a function of incident energy ($E/\Delta_0$) for different values of $\chi$ with $\alpha =0$ and $\alpha =1$. 
It is clearly visible that for $\alpha = 0$ (graphene) and $U_0 =0$ (Fig.\ref{Fig4}a), the tunneling conductance at the gap edge reaches a value close to $2G_0$ when $\chi = (n + 1/2)\pi$ (for any integer $n$), the conductance $G$ repeats for $\chi=n\pi$. While for $\alpha =1$ (dice) and $U_0 =0$ (Fig.\ref{Fig4}b), the tunneling conductance at the gap edge reaches a value close to $2G_0$ when $\chi = n\pi$ with a periodicity $\pi$. However, with $U_0 \gg E_F$ for both the graphene (Fig.\ref{Fig4}c) and the dice (Fig.\ref{Fig4}d) cases, the maximum conductance occurs at $\chi=n\pi/2$ with a periodicity $\pi/2$.\\

\begin{figure}
	\centering
	\includegraphics[width=0.46\textwidth]{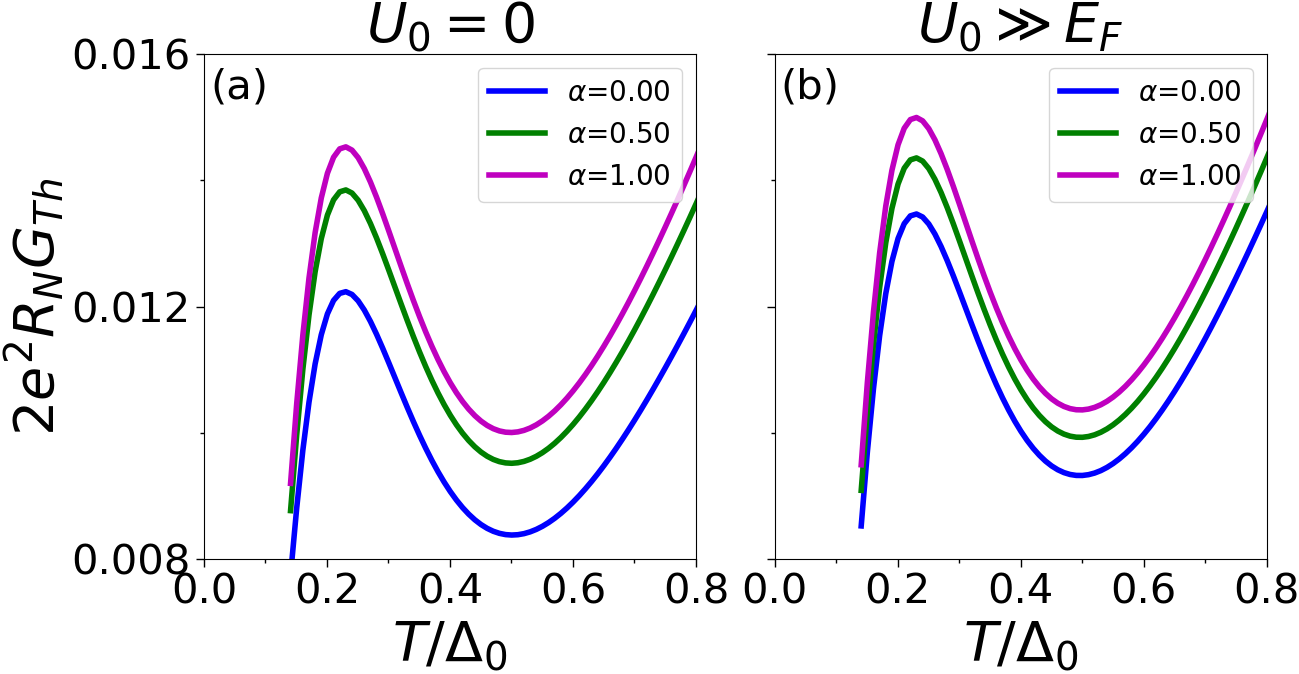}
	\caption{The variation of the dimensionless thermal conductance, $2e^2R_NG_{Th}$ as a function of temperature (scaled by $\Delta_0$), for different values of $\alpha$ (a) for $U_0 = 0$ and (b) for $U_0 \gg E_F$.}
	\label{Fig7}
\end{figure}

\begin{figure}
\centering
\includegraphics[width=0.47\textwidth]{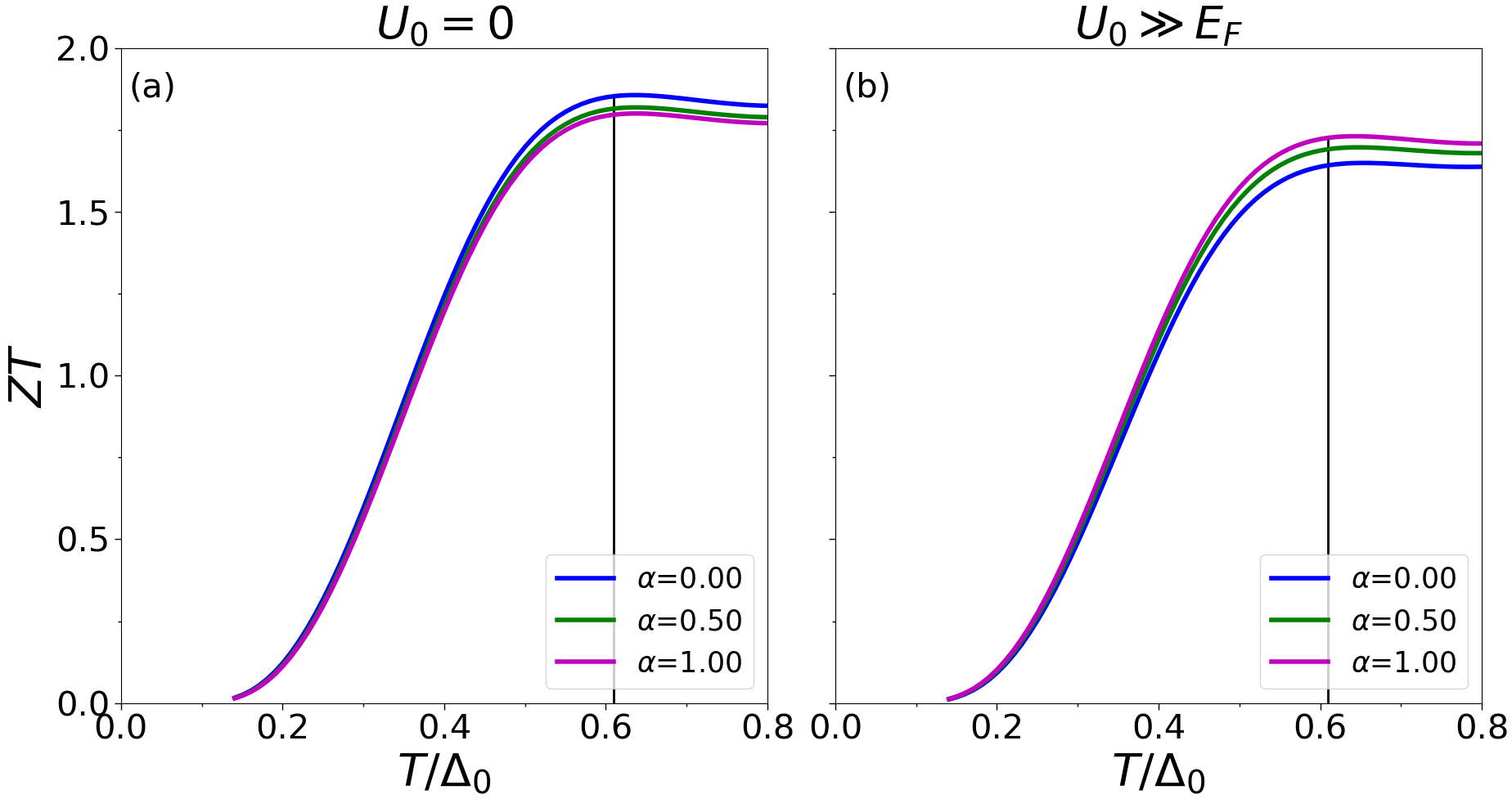}
	\caption{The variation of Figure of Merit, $ZT$ as a function of temperature (scaled by $\Delta_0$), for different values of $\alpha$ (a) for $U_0 = 0$ and (b) for $U_0 \gg E_F$..}
	\label{Fig8}
\end{figure}

\textbf{\textit{B. Seebeck Coefficient}:} 
The variation of the Seebeck coefficient $S$ as a function of temperature (in units of superconducting gap, $\Delta_0$) is shown in the Fig.\ref{Fig5}a, and Fig.\ref{Fig5}b. Here also, we consider two cases: $(i)$ $U_0=0$ and $(ii)$ $U_0\gg E_F$ with $E_F^N=50\Delta_0$. Further, we consider a particular value of $\chi$, where the charge conductance records the maximum value. We represent the Seebeck coefficient $S$ in unit of ($k_B/e$). For both the cases $(i)$ (see Fig.\ref{Fig5}a), and $(ii)$ (see Fig.\ref{Fig5}b), the Seebeck coefficient shows maximum values at $T\sim0.3\Delta_0$.  It is understood that the Seebeck coefficient increases initially with temperature and after attaining a certain value it decreases for all values of $\alpha$.

Further, we observe that the maximum of the Seebeck coefficient decreases with the increasing strength of $\alpha$ for $U_0=0$ (Fig.\ref{Fig5}a), while for $U_0 \gg E_F$, the reverse trend  is obtained in Fig.\ref{Fig5}b.\\

\textbf{\textit{C. Charge Conductance}:} 
Furthermore, from Eq.(\ref{eq25}) we can see that “figure of merit” $ZT$, which defines the efficiency of this system as a thermopower device, depends upon the Seebeck coefficient $S$, the charge conductance $G_c$, and the thermal conductance $G_{Th}$. To understand the effect of charge and thermal conductance on $ZT$, we show the results of the charge and the thermal conductance of the $\alpha-T_3$ based NIS junction. In Fig.\ref{Fig6} we present the variation of a dimensionless quantity, $2eR_NG_c$ as a function of temperature (scaled by $\Delta_0$) for a fixed value of $\chi$ for two different cases $(i)$ $U_0 = 0$ (see Fig.\ref{Fig6}a), and $(ii)$ $U_0 \gg E_F$ (see Fig.\ref{Fig6}b). Here $G_c$ increases monotonically with an increase in $T/\Delta_0$ and after attaining a certain values it saturates for all values of $\alpha$ for both the cases. Unlike the Seebeck coefficient, the maxima of the charge conductance increases with the increase of $\alpha$ for both the cases.\\

\textbf{\textit{D. Thermal Conductance}:}
In Fig.\ref{Fig7} we present the variation of a dimensionless quantity, $2e^2R_NG_{Th}$ as a function of temperature (scaled by $\Delta_0$) for a fixed value of $\chi$ for two different cases $(i)$ $U_0 = 0$ (see Fig.\ref{Fig7}a), and $(ii)$ $U_0 \gg E_F$ (see Fig.\ref{Fig7}b). Here $G_{Th}$ increases rapidly with an increase of $T/\Delta_0$ in the low temperature regime. After attaining a maximum value (where the Seebeck coefficient attains maxima), it decreases, and after attaining a minimum value (where the charge conductance gets maxima), it again increases with temperature for all values of $\alpha$ for both the cases. Here, unlike the Seebeck coefficient, the maxima of thermal conductance increases with the increase of $\alpha$ for both the cases.\\

\textbf{\textit{E. Figure of Merit}:}
Now the variation of the figure of merit, $ZT$ as a function of temperature (in units of the superconducting gap, $\Delta_0$) is shown in the Fig.\ref{Fig8}. It shows that initially with temperature the efficiency of the system increases,  attains a maximum value at $T\sim 0.62\Delta_0$ and then decreases. For the Fermi surface matched condition, that is, $U_0 =0$ (Fig.\ref{Fig8}a) the maximum efficiency of the NIS junction device decreases as we increase the strength of the $\alpha$ parameter from $0$ to $1$. Further, for the Fermi surface mismatch condition i.e., $U_0 \gg E_F$ (Fig.\ref{Fig8}b), the maximum efficiency increases as we increase the value of $\alpha$. Thus, for two different cases, the figure of merit follows the same trend as that of the Seebeck coefficient as a function of $\alpha$.\\


\begin{figure}
\includegraphics[width=0.45\textwidth]{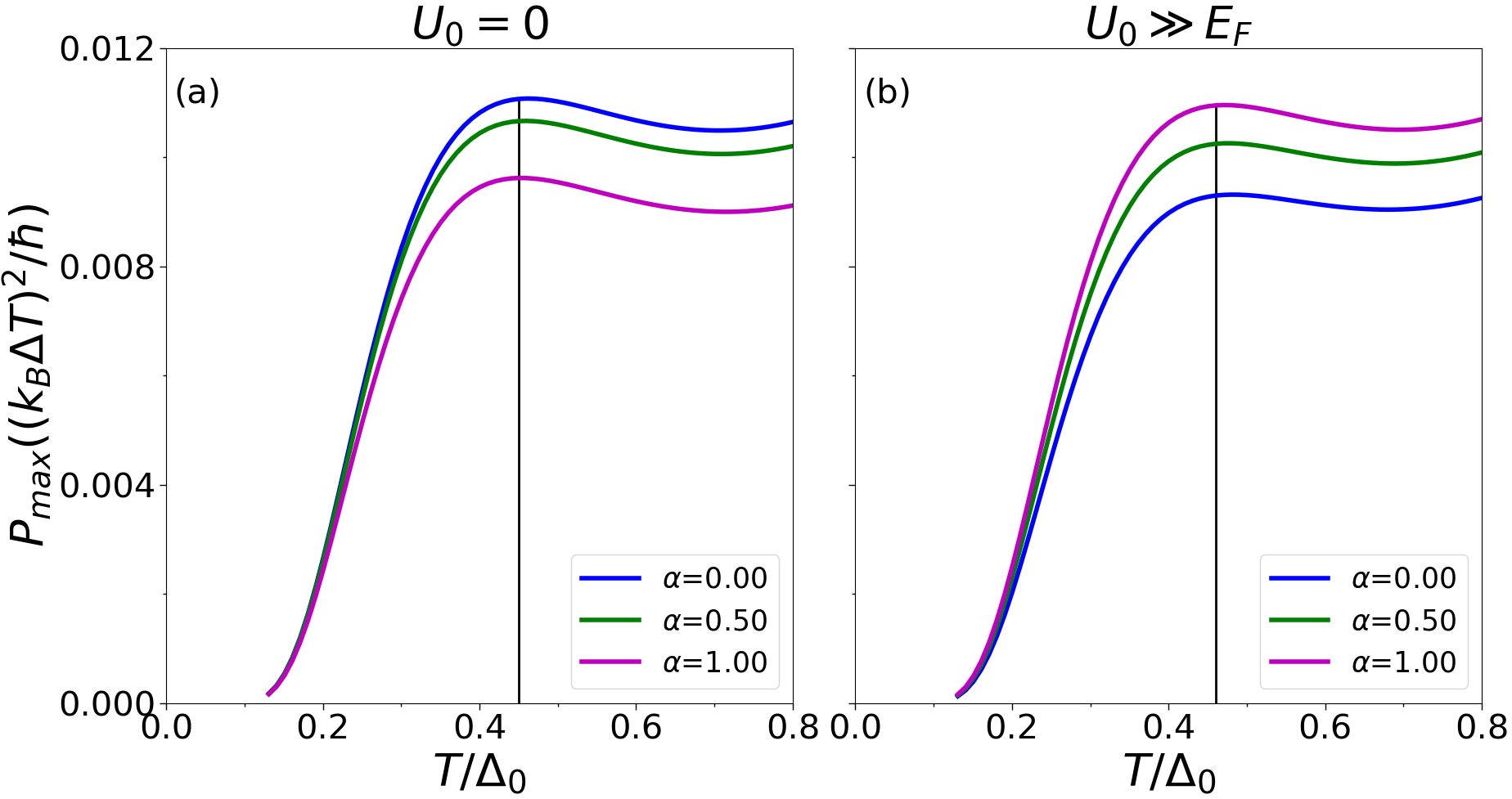}
\caption{Plots of the maximum power, $P_{max}$ [in unit of $(k_B\Delta T)^2/\hbar)$] as a function of temperature (scaled by $\Delta_0$), for different values of $\alpha$ (a) for $U_0 = 0$ and (b) for $U_0 \gg E_F$.}
\label{Fig9}
\end{figure}

\begin{figure}
	\centering
	\includegraphics[width=0.47\textwidth]{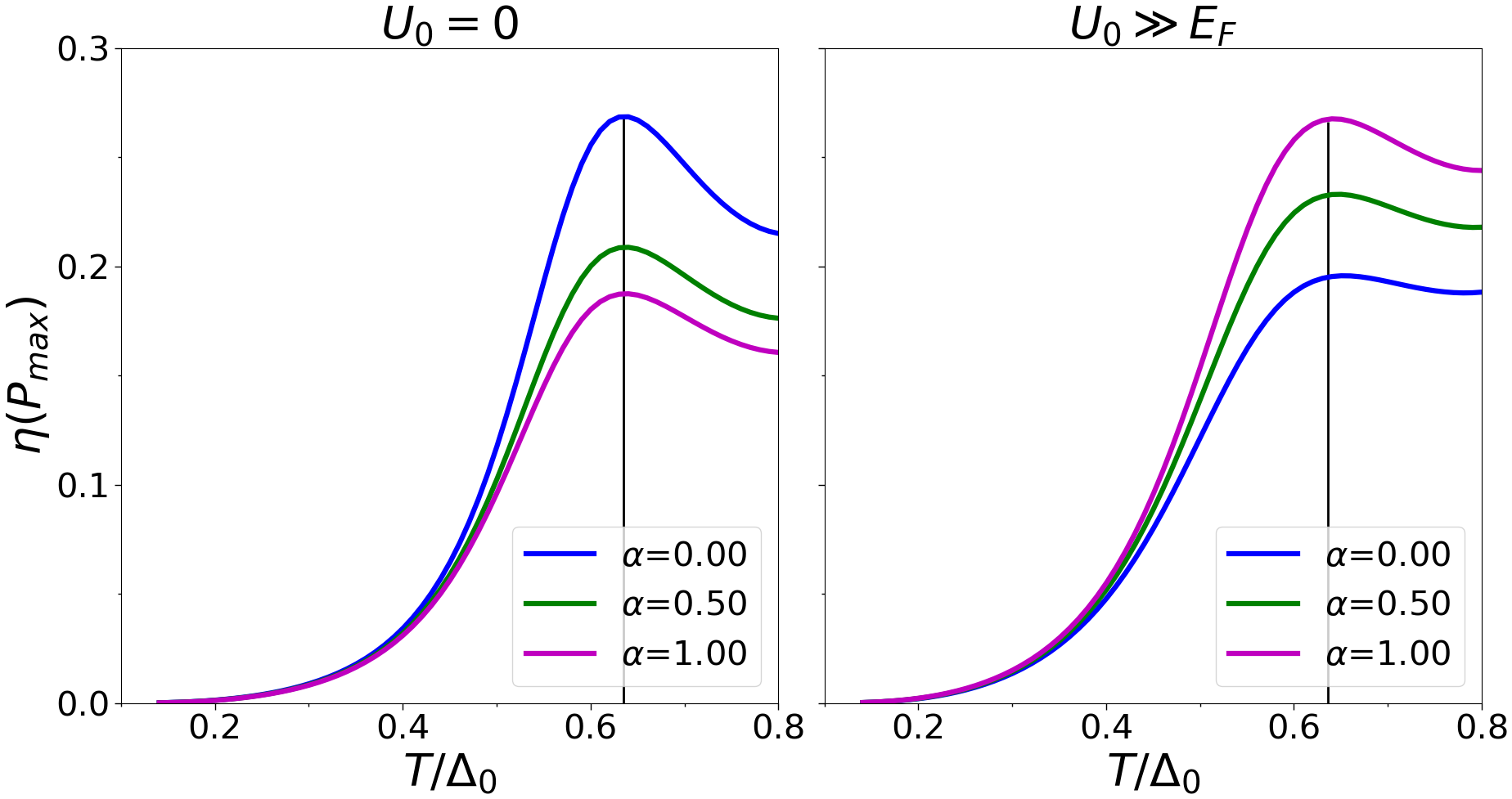}
	\caption{ Plots of the efficiency at maximum power, $\eta$ $(P_{max})$ [in unit of $\eta_c$] as a function of temperature (scaled by $\Delta_0$), for different values of $\alpha$ (a) for $U_0 = 0$ and (b) for $U_0 \gg E_F$.}
	\label{Fig10}
\end{figure}

\textbf{\textit{F. Maximum Power}:}
 The maximum power $P_{max}$ is plotted as a function of temperature in Fig.\ref{Fig9}. As earlier, here also we consider two cases: $(i)$ for $U_0=0$ (Fig.\ref{Fig9}a), and $(ii)$ for $U_0 \gg E_F$ (Fig.\ref{Fig9}b). The maximum power follows the same trend as that of the Seebeck coefficient. We can see that the maximum power is obtained where the Seebeck coefficients get their maximum value irrespective of the $\alpha$ value. The maximum power is obtained at the temperature ($T/\Delta_0$) $\sim 0.47$ for all two cases mentioned above and for all values of the parameter $\alpha$. As earlier, we observe that the maximum power decreases with the increasing strength of $\alpha$ for $U_0 = 0$, while for $U_0 \gg E_F$, the reverse trend is obtained.\\
 
\textbf{\textit{G. Efficiency at Maximum Power}:}
From Eq.(\ref{eq24}) we can see that the maximum efficiency $\eta(P_{max})$ of the system depends upon the figure of merit $ZT$ of the system. In Fig.\ref{Fig10} we present the efficiency for two pre-defined cases (see Fig.\ref{Fig10}a and Fig.\ref{Fig10}b). The efficiency when the system operates at the maximum power follows the same trend as that of the figure of merit $ZT$ and the Seebeck coefficient $S$ as $\alpha$ is varied. Thus, the best efficiency is obtained whenever the temperature ($T/\Delta_0$) is $\sim 0.62$ for the two different cases and all values of parameter $\alpha$.\\
 
\begin{figure}
	\centering
	\includegraphics[width=0.47\textwidth]{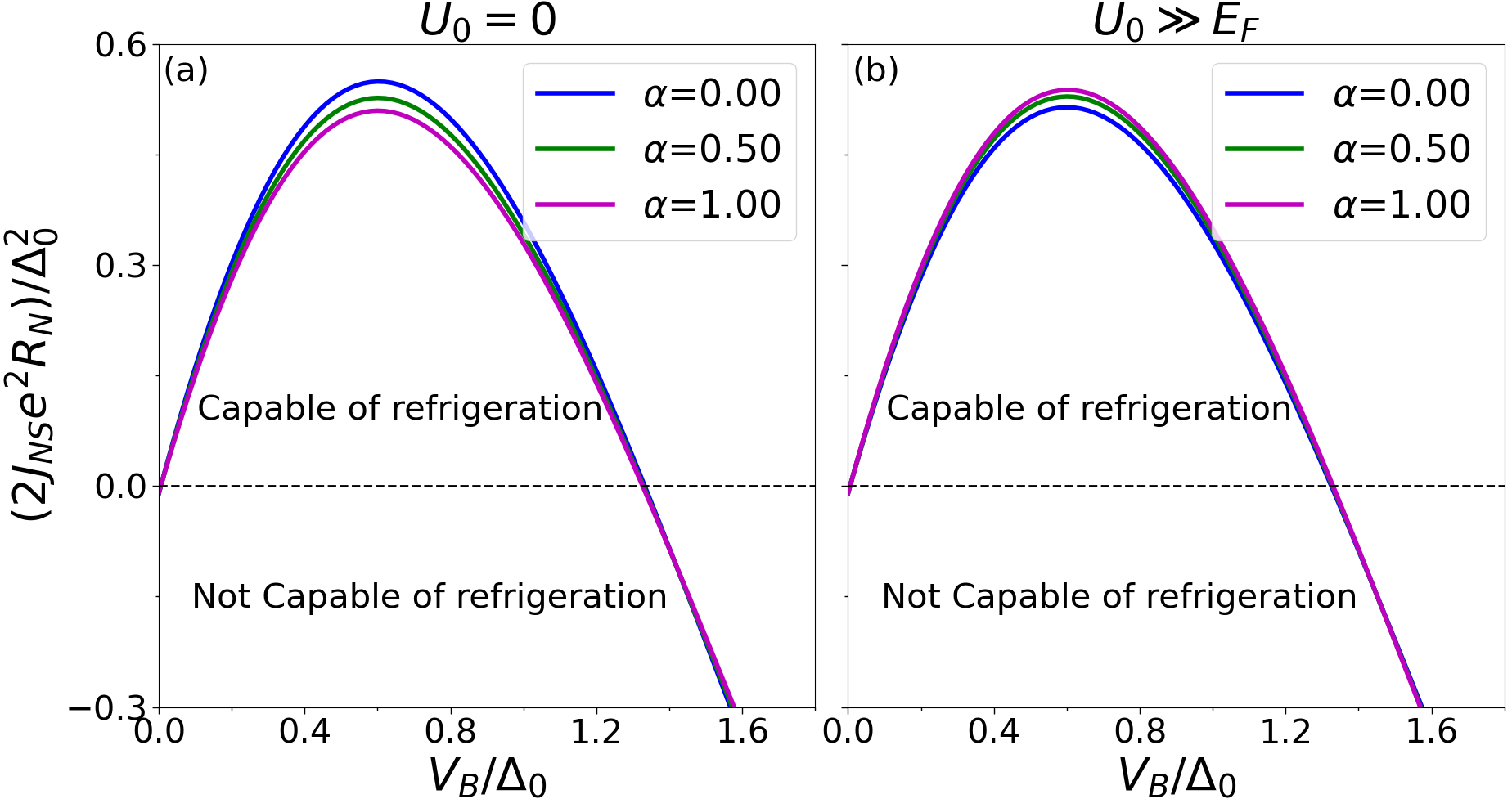}
	\caption{The variation of the dimensionless thermal current, $(2J_{NS}e^2R_N)/\Delta_0^2$ as a function of biasing voltage $V_B$ (scaled by $\Delta_0$) for different values of $\alpha$. (a) For $U_0=0$, and (b) for $U_0 \gg E_F$. The operating regions are indicated.}
	\label{Fig11}
\end{figure}

\textbf{\textit{H. Thermoelectric Cooling}:}
Here, we show the results of the thermoelectric cooling of the $\alpha-T_3$ based NIS junction. Fig.\ref{Fig11} presents the variation of a dimensionless quantity $2J_{NS}e^2R_N/\Delta_0^2$ ($J_{NS}$ being the thermal current) as a function of the biasing voltage $V_B$ (in units of the superconducting gap $\Delta_0$), for a fixed $\chi$ where the temperature is fixed at $T = 0.5\Delta_0$. It is observed that at zero bias voltage, the rate of the thermal current extracted from the cold (normal) reservoir is negative. To achieve cooling effects, a lower threshold voltage of the battery, namely, $V_{lower}$ is needed. Also beyond an upper threshold voltage $V_{upper}$, the refrigeration effect does not exist. For a very small range of the biasing voltage ($\sim 0.005\Delta_0$-$1.3\Delta_0$), the thermoelectric cooling process is efficient and the maximum refrigeration occurs at around, $V_B \sim 0.6\Delta_0$ for all values of $\alpha$. Further, the maximum refrigeration decreases as we increase the strength of the $\alpha$ parameter for $U_0 =0$ case (Fig. \ref{Fig11}a). Furthermore, for the Fermi surface mismatch condition i.e., for $U_0 \gg E_F$, (Fig. \ref{Fig11}b) the maximum refrigeration increases as we increase the strength of the $\alpha$ parameter.
\\

\section{Conclusions}
\label{Sec4}
In conclusion, we discuss the tunneling conductance of $\alpha-T_3$ based NIS junctions. We have demonstrated that the tunneling conductance exhibits periodic oscillatory behavior as a function of the barrier strength of the junction. The periodicity and the amplitudes of the oscillations depend on $\alpha$ and as well as on the values of $U_0$. We have investigated the Seebeck coefficient, the thermoelectric figure of merit, maximum power, and the efficiency at maximum power of this junction device and explained how $\alpha$ plays a role in determining these properties. Further, the thermoelectric cooling of this junction device have been studied in details. We have ascertained whether a graphene or a dice lattice is more suitable for a thermoelectric device, and the answer depends on the superconducting potential $U_0$. We observe that for $U_0=0$, graphene ($\alpha=0$) is more feasible for constructing a thermoelectric device, whereas for $U_0 \gg E_F$, the dice lattice ($\alpha=1$) is more suitable candidate. Since the effective barrier potential ($\chi$), the superconducting potential $U_0$, and the hopping strength $\alpha$ can be tuned by external means, our study on the conductance and thermoelectric properties can be important inputs to the experimental studies that aim to provide the desired conductance and thermoelectric properties of $\alpha-T_3$ based NIS junction devices.\\

\textbf{Acknowledgements}

MI and PK gratefully acknowledge Prof. S. Basu for useful discussions.

\end{document}